%% file: 00_main.tex
\documentclass[acmtog]{acmart}

\def\BibTeX{{\rm B\kern-.05em{\sc i\kern-.025em b}\kern-.08emT\kern-.1667em\lower.7ex\hbox{E}\kern-.125emX}}
    
\setcopyright{none}
\usepackage{amssymb}
\usepackage{xcolor}
\usepackage{wrapfig}

\renewcommand{\eqref}[1]{Eqn.~\ref{eq:#1}}

\begin{document}

\newcommand{\fig}[1]{Fig.~\ref{#1}}
\newcommand{\sectnum} [1] {\ref{#1}}
\newcommand{\sect} [1] {Sec.~\sectnum{#1}}

%
% The "title" command has an optional parameter, allowing the author to define a "short title" to be used in page headers.
\title{\emph{DecoSurf}: Recursive Geodesic Patterns on Triangle Meshes}

%
% The "author" command and its associated commands are used to define the authors and their affiliations.
% Of note is the shared affiliation of the first two authors, and the "authornote" and "authornotemark" commands
% used to denote shared contribution to the research.
\author{Giacomo Nazzaro}
\affiliation{%
  \institution{Sapienza University of Rome}
%  \streetaddress{1 Th{\o}rv{\"a}ld Circle}
  \city{Rome}
  \country{Italy}}
\email{nazzaro@di.uniroma1.it}

\author{Enrico Puppo}
\affiliation{%
  \institution{University of Genoa}
  \city{Genoa}
  \country{Italy}
}

\author{Fabio Pellacini}
\affiliation{%
 \institution{Sapienza University of Rome}
 \city{Rome}
% \state{}
 \country{Italy}}
 
%
% By default, the full list of authors will be used in the page headers. Often, this list is too long, and will overlap
% other information printed in the page headers. This command allows the author to define a more concise list
% of authors' names for this purpose.
\renewcommand{\shortauthors}{Nazzaro, et al.}

\input{00_abstract}

%CCS


%keywords

%
% Keywords. The author(s) should pick words that accurately describe the work being
% presented. Separate the keywords with commas.
% \keywords{datasets, neural networks, gaze detection, text tagging}

\begin{teaserfigure}
  \centering
  \includegraphics[width=0.97\textwidth]{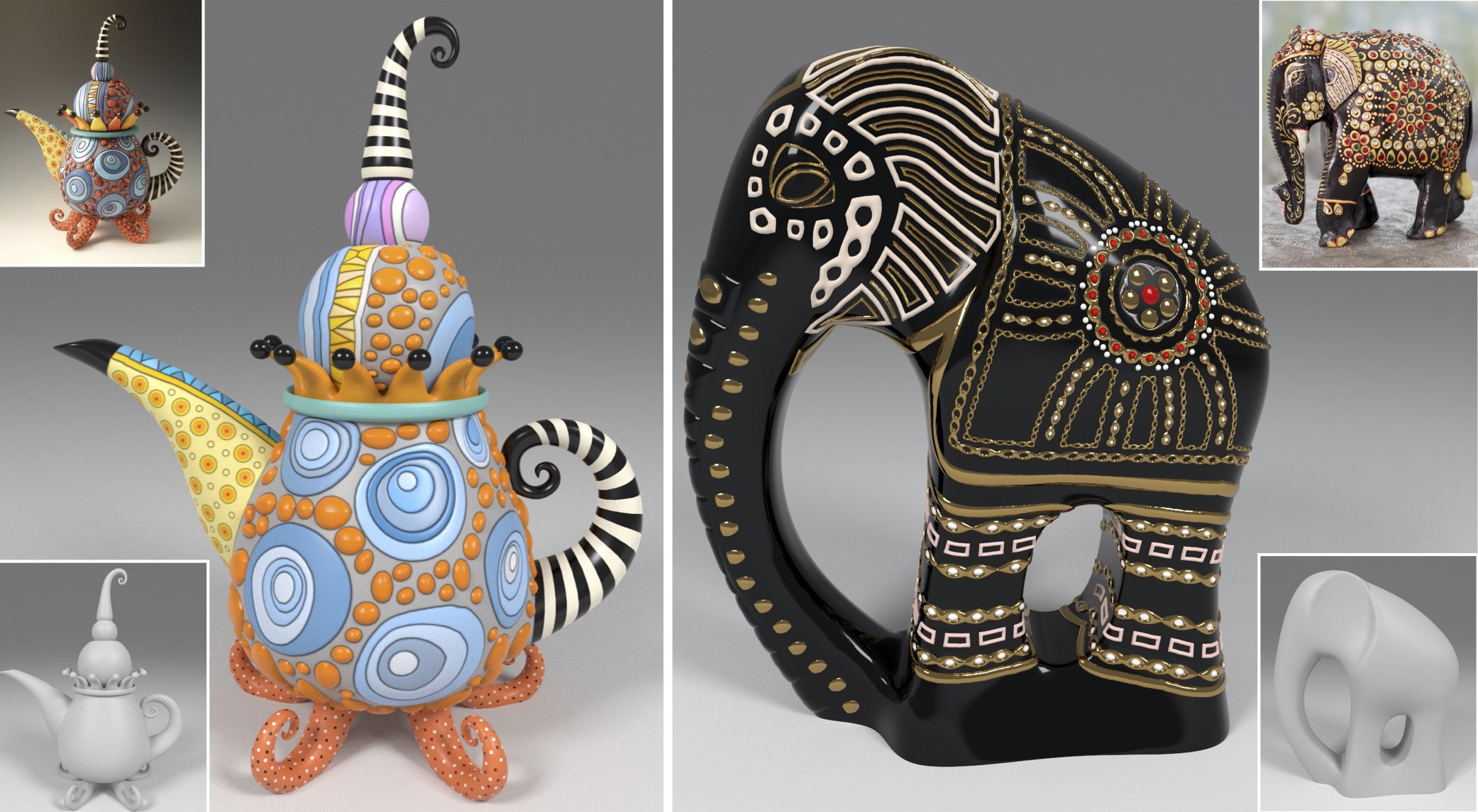}
  \caption{Models decorated with \emph{DecoSurf} (large images) to match the style of real-world 
  photographs (upper insets), together with the input meshes %undecorated models 
  (lower insets: teapot 1.5M triangles, elephant 2M triangles). 
  We model decorations by recursively splitting a surface into progressively 
  finer regions, to which we apply material and displacement variations. 
  All patterns were constructed with only four operators
  that split regions along the isolines or integral curves of scalar fields derived from geodesic computations. 
  Photos of real objects: teapot by Natalya Sots; elephant by  Dayal J. Daryanani. 
}
  \label{fig:teaser}
  \Description{Image}
\end{teaserfigure}

\maketitle

\input{01_introduction}
\input{02_related}
\input{03_patterns}
\input{04_implementation}
\input{05_results}
\input{06_conclusions}

%
% The next two lines define the bibliography style to be used, and the bibliography file.
\bibliographystyle{ACM-Reference-Format}
\bibliography{references}

\end{document}

%% file: 00_abstract.tex
% !TEX root = decosurf.tex

\begin{abstract}
\label{sec:abstract}
%Many artisanal objects are decorated with simple shapes
%forming complex patterns. In this paper we show that many such patterns 
In this paper, we show that many complex patterns, 
which characterize the decorative style of many artisanal objects, 
can be generated by the recursive application of only four operators.
Each operator is derived from tracing the isolines or the integral curves 
of geodesics fields generated from selected seeds on the surface.
Based on this formulation, we present an interactive application that lets 
designers model complex recursive patterns directly on the object surface,
without relying on parametrization.
We support interaction on commodity hardware on meshes of a few million triangles, by combining 
light data structures together with an efficient approximate graph-based geodesic solver.
We validate our approach by matching decoration styles from real-world photos, 
by analyzing the speed and accuracy of our geodesic solver, 
and by validating the interface with a user study.
\end{abstract}

%% file: 01_introduction.tex
% !TEX root = decosurf.tex

\section{Introduction}
\label{sec:intro}

Many artisanal objects are decorated with complex patterns, consisting of simple 
basic shapes, like dots, lines and circles organized into nested hierarchies. 
The complexity of these decorations comes from repetitive structures that appear 
handmade instead of being arranged geometrically. 
Artisans create such decorative patterns by recursively splitting regions to 
add progressively finer details. 
\fig{fig:teaser} shows examples of these \emph{recursive decorative patterns} 
that are often found in handcrafted Italian ceramics and wood pieces from Indian art. 
Note how surfaces are decorated both by changing color and material, 
but also by altering the surface locally, e.g., adding or removing clay in ceramics. 

With the advent of 3D printing and the need for personalization, 
digital modeling of recursive decorative patterns becomes important.
Reproducing these patterns with digital sculpting tools may require
hours of work by a skilled artist to decorate a single object.
The alternative would be to use procedural programs, grammars or node graphs, 
but such techniques are notoriously hard for non-technical artists and novices. 
In the research literature, we are not aware of any work specifically 
addressing this class of patterns. 

In this paper, we present \emph{DecoSurf}, a general yet simple framework for 
modeling recursive decorative patterns. 
\fig{fig:teaser} and \fig{fig:results} show examples of 
decorations created with our model that match real-world styles.
We obtain our patterns as recursive applications of only four generic operations,
each of which splits a region of the surface into smaller regions to generate 
progressively finer decorations. 
At each split, we optionally apply material variations and displacement to 
define the final look. 
The key insight of our work is that a large class of decorative patterns can be 
modeled by recursively cutting regions along the isolines and the integral curves 
of geodesic fields computed from appropriately selected seeds, namely the 
regions' borders or points sampled within each region. 
To better integrate our work with designers workflow, 
we focus on controlling pattern generation using an interactive application,
rather than asking artists to write procedural programs or grammars.
To reduce repetitive work though, we also support easy-to-use predefined 
procedural patterns that are integrated into the interactive workflow.

We model patterns directly on surfaces, relying on geodesic distances for three 
main reasons. 
First, we want to work directly on the manifold using its intrinsic metric, 
thus avoiding distortions, discontinuities and topological artifacts, 
which arise from parametrization and require complex handling during 
pattern synthesis. 
Second, the geodesic metric encodes many intrinsic properties of the surface 
and remains well defined after mesh split, even when applying displacement. 
This in turn allows us to define a closed algebra of operators that can be 
applied at will recursively to generate a remarkable variety of decoration styles. 
Finally, having one metric for all patterns is practical to implement and 
easier to optimize compared to using a variety of unrelated operators.  

We address surfaces represented with fine triangle meshes, similarly to 
digital sculpting, using roughly a million triangles per surface to ensure that 
patterns are finely represented and to model displacement precisely. 
The main challenge is thus to support interactivity for all operations involved 
in pattern generation. 
A key point of this work is the careful design of our data structures, which 
allow us to implement simple algorithms that scale efficiently to highly 
tessellated meshes.
Our patterns depend on two main operations: computation of geodesic distance 
fields and cutting triangle meshes along isolines and integral curves of these 
fields. 
Mesh cutting is relatively straightforward to implement efficiently, 
while geodesic computations are notoriously expensive to perform. 
To address this problem we combine data structures with a small footprint 
together with a fast, graph-based, approximate, geodesic solver that 
allows for dynamic update and
scales very well with mesh complexity. 

We have implemented a prototype user interface, which exposes the pattern
operators directly. \fig{fig:interface} shows the interface, while the 
supplemental video demonstrates the use of each operator and the creation of a 
complex decoration. While we focus on interactive editing, we also show that our 
operators can be used procedurally. 

We validate how well \emph{DecoSurf} models real-world recursive decorations 
in three manners. 
First, we model decorations to match the styles of real-world artisanal objects. 
We found that we can replicate complex patterns from different styles with ease. 
Second, we measure the accuracy and speed of our geodesic solver and of the overall 
editor to show that we can maintain both accuracy and interactivity while modeling. 
Third, we run a user study where we ask subjects to model complex patterns with 
our interface and whether those patterns match real-world surfaces. 
The main results of the study are that users can reproduce complex patterns with 
ease and they find that our patterns match real-world styles. 

In summary, all these results were made possible by a few key technical contributions:
\begin{itemize}
\item the definition of a small set of operators, based on computing geodesic
fields from appropriate seeds, which can be combined recursively to obtain 
many intricate decorations;
\item the implementation of a fast geodesic solver that is interactive on meshes 
with millions of triangles, requires no expensive pre-computation, can be 
updated efficiently upon mesh editing, and is accurate enough at high triangle 
count;
\item the implementation of a user interface to generate recursive decorative 
patterns that is simple-to-use while allowing the creation of complex decorations;
\item the extension of our model to a procedural use;
\item the validation of our model by matching real-world photos, testing accuracy 
and speed, and running a user study.
\end{itemize}

%% file: 02_related.tex
% !TEX root = decosurf.tex

\section{Related work}
\label{sec:related}

\paragraph{Procedural Patterns}
Complex patterns may be generated with procedural methods, for which 
there exists significant literature. The main differences between 
procedural algorithms depend on the domain in which they perform the synthesis, 
the control given to designers, and the underlying algorithmic formulation.

Procedural texturing is at the base of most pattern synthesis used both in the 
literature and in practice. \cite{Ebert:2002} present the basic methods for 
procedural texturing, which are applied commonly in practical applications.
These methods are very general in the patterns that they define, but they all
require the user to describe the pattern using either code or visual languages, 
and pattern generation requires evaluating a function at each point in either 
2D textures, or 3D volumes. 
We differ substantially from this approach in that we generate structured patterns 
directly on objects' surfaces instead of relying on generating textures that 
would require a parametrization. 
Also, recursive patterns cannot be constructed by evaluating a function on each 
surface location but they require recursive evaluation of the entire pattern.

A second class of methods generates new patterns directly from example images
via non-parametric texture synthesis implemented as variations of constrained 
random sampling, see \cite{Wei:2009} for a survey, or via neural network and 
generative adversarial networks, e.g., \cite{Sendik:2017,Zhou:2018}. 
While these methods are remarkably reliable for unstructured patterns, 
they often fail to capture patterns with complex structural properties. 
Structured recursive patterns, in particular, cannot be captured by 
non-parametric synthesis as shown in \cite{Santoni:2016} for the simpler 
2D case.

The last class of methods, and the one more closely related to our work, is
based on stochastic grammars that recursively split shapes into smaller 
components. \cite{Santoni:2016} show that group grammars can be used to 
describe tangle patterns in the 2D domain.  Group grammars are extensions of 
shape grammars popularized for modeling buildings, see \cite{Schwarz:2015} 
for a recent review. Groups grammars have been extended to procedural animation 
as motion grammars shown in \cite{Carra:2019}. 
\cite{Li:2011} uses grammar guided by vector fields to place external details 
on surfaces, but cannot handle recursive patterns. 
An alternative approach to grammars is to use a custom programming 
language to express stationary discrete textures, as shown in \cite{Loi:2017}. 
Of these works, \cite{Santoni:2016} is closest to ours since it also models 
recursive patterns. The main differences between the two approaches are that 
they use ad-hoc operators only suitable to tangle art in 2D, while we propose 
general operators working on 3D surfaces. Also, we focus on interactive editing, 
versus building a grammar system. In fact, our method may be considered a 
super-set of theirs, while none of the images in this paper could have been 
generated using the formulation in \cite{Santoni:2016}.

\paragraph{Geodesics}
Broadly speaking, the literature offers three classes of methods
for computing geodesic distance fields and paths over surface meshes. 
A survey can be found in \cite{Bose:2011}, although several more recent 
methods exist. 

Exact methods for polyhedral surfaces stem from the seminal algorithms 
presented in \cite{Mitchell:1987} and \cite{Chen:1990}, which were improved 
several times in the literature. But even the most recent methods in this line 
\cite{Qin:2016} are too slow to support interaction on moderately large meshes. 

Graph-based methods provide approximated solutions of polyhedral geodesics, 
by restricting possible paths to chains of arcs in a graph.
A straightforward shortest path computation on the network of edges results in
poor distance estimation and wiggly paths. In \cite{Campen:2013}, such paths 
are improved by computing shortcuts on-the-fly. 
\cite{Lanthier:1997bg,Lanthier:2001} precompute an extended graph adding 
Steiner nodes on edges to improve the approximation.
Most recent methods in this class \cite{Wang:2017,Ying:2013} precompute a 
sub-graph of the complete graph connecting all vertices, such that each possible 
geodesic path can be approximated well with a sequence of paths in such graph.
We rely on a simple variation of graph-based methods, discussed in 
\sect{sub:graph}, which scales well to large meshes, and can be maintained 
very easily and efficiently upon mesh refinement.

PDE methods define the geodesic distance problem in terms of partial 
differential equations \cite{Kimmel:1998,Crane:2013} and provide results that 
approximate geodesic distances on a smoothed surface. 
The Fast Marching Method \cite{Kimmel:1998} requires no pre-processing and could 
be adapted well to local computations and dynamic mesh refinement, but it is overly
slow for our needs. 
A parallel version of FMM has been proposed very recently \cite{ROMEROCALLA2019}, 
which greatly improves time performances, yet remaining slower than our approach, with a comparable accuracy. 
The heat method \cite{Crane:2013} requires resolving sparse linear systems of the size of the mesh. 
It can run very efficiently on relatively large meshes by exploiting 
pre-factorization, but it cannot be extended easily to manage local computations, 
or dynamic mesh refinement.
The former issue can be partially addressed by incorporating the methods 
proposed in \cite{Herholz:2017,Herholz:2018}.
%We considered using these approaches in our work, but found that the need
%for pre-factorization hindered our ability to quickly slice and displace the mesh. 

%% file: 03_patterns.tex
% !TEX root = decosurf.tex

\begin{figure}[t]
\centering
\includegraphics[width=\linewidth]{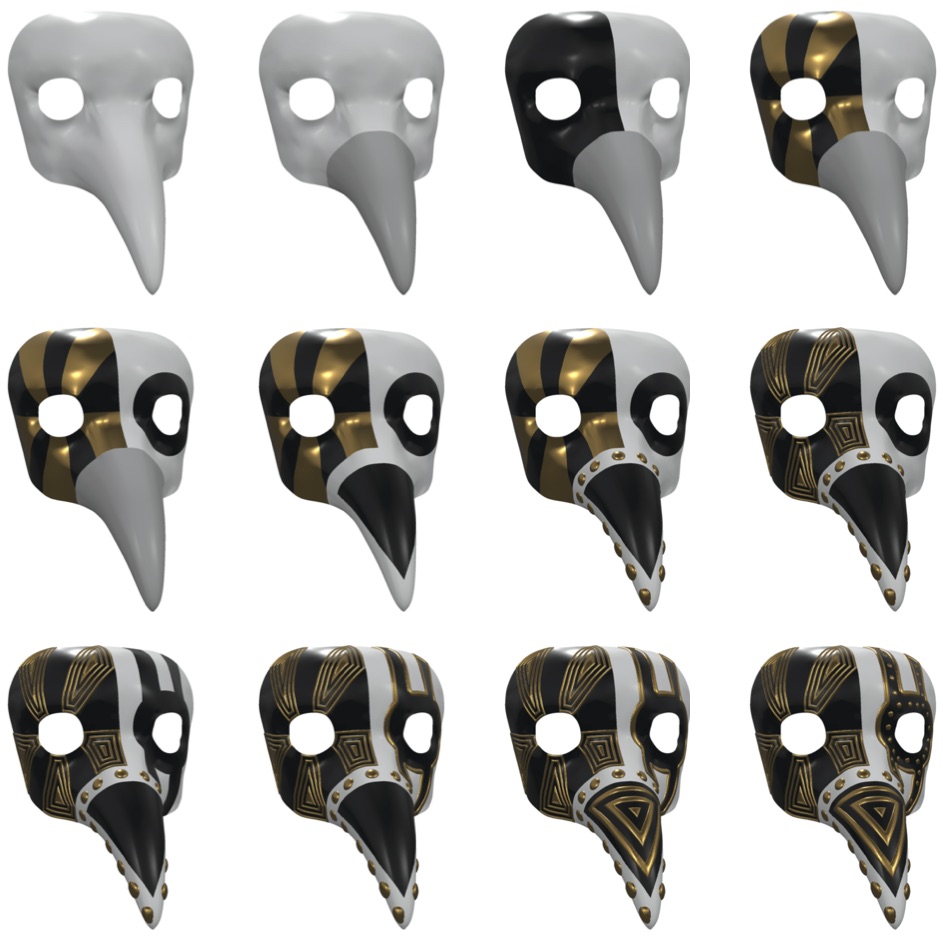}
\caption{Complex recursive patterns are obtained in just eleven steps 
(left to right, top to bottom) starting at a base shape (upper left). 
%(first row, left to right) basic shape, polyline, polyline, stream; (second row) twice contour from boundary, contour from sampled points, multiple contours from boundary; (third row) multiple contour from boundary edge, contour from boundary, multiple contours from boundary, contour from sampled points. 
At each step we apply a basic operation, which recursively splits the surface into finer decorations.
The generated patterns naturally follow the shape of the surface, as they are based on geodesic distance fields.}
\label{fig:sequence}
\Description{Image}
\end{figure}

\begin{figure*}[t]
\centering
\includegraphics[width=0.95\textwidth]{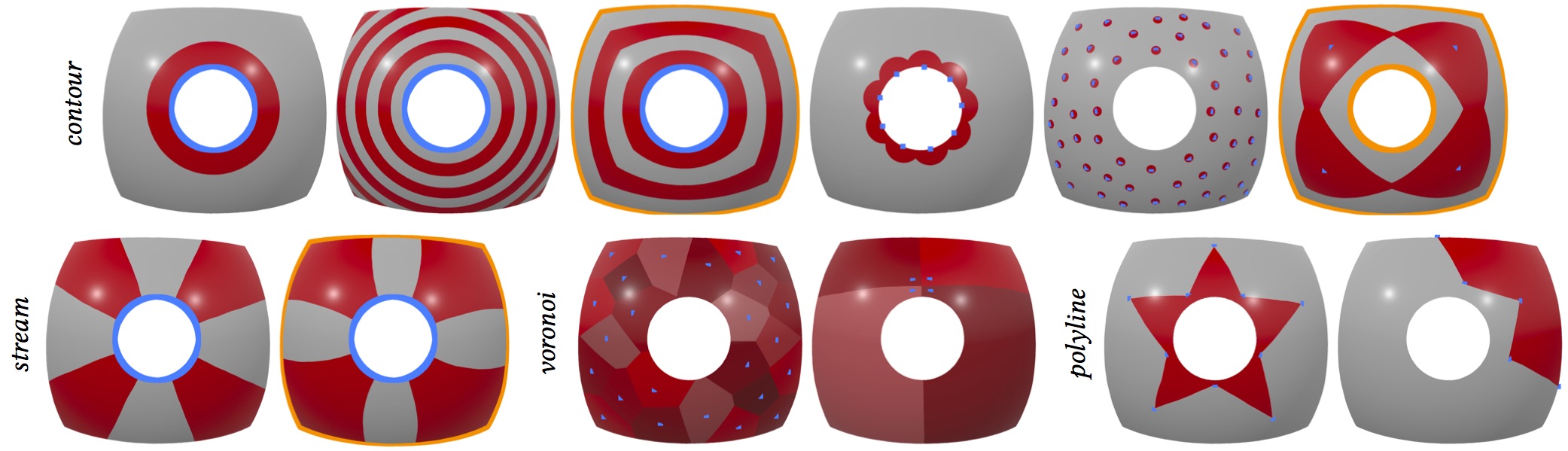}
\caption{Example decorations created by applying only one operator with 
different field and seed sets.
(Top, left to right) \emph{Contour} operator with 
(1) \emph{dist} field from the inner boundary (in blue) cutting along a single isovalue;  
(2) \emph{dist} field from the inner boundary with multiple isovalues;
(3) \emph{blend} field between inner (blue) and outer (orange) boundary, cutting along multiple isovalues;
(4) \emph{dist} field from uniform sampling of points along the inner boundary;
(5) \emph{dist} field from Poisson sampling of points on the surface;
(6) \emph{blend} field between sampled points (blue) and boundaries (orange).
(Bottom left, left to right) \emph{Stream} operator with 
(7) \emph{dist} field from inner boundary; 
(8) \emph{blend} field between inner and outer boundary.
(Bottom middle, left to right) \emph{Voronoi} operator with  
(9) \emph{dist} field from points selected with Poisson sampling; 
(10) \emph{dist} field from manually selected points.
(Bottom right, left to right) \emph{Polyline} operator with
(11) closed polyline on a multiply connected region;
(12) open polyline connecting two points on the same boundary loop.}
\label{fig:operators}
\Description{Image}
\end{figure*}

\section{Recursive Geodesic Patterns}
\label{sec:operators}

We aim at producing patterns defined as recursive subdivisions of a surface %2-manifold
%, possibly with boundary, 
into nested shapes. % on the surface.
We show that a small arsenal of simple operators, defined upon geodesic fields 
and combined in the context of a coherent framework, 
together with the recursive nature of decorative patterns,
is sufficient to generate a great variety of patterns that 
match the style of handmade decorations.
\fig{fig:sequence} shows an example of a pattern generation sequence.

\subsection{Overview}

\paragraph{Pattern Representation}
%We use the following terminology to describe our work.
Let $\mathcal M$ be a 2-manifold, possibly with boundary.
A \emph{region} $R$ is a connected subset of 
$\mathcal M$ bounded by a finite number of oriented boundary loops.
A \emph{recursive pattern} $\mathcal T$ is a tree of regions,
having its root at $\mathcal M$ and such 
that the children of each region $R$ in $\mathcal T$ form a partition of $R$. 
%It follows that t
The leaves of $\mathcal T$ form a partition of $\mathcal M$
and those regions ultimately define the pattern.

\paragraph{Operators}
We construct decorative patterns by recursively applying a small set of four 
operators that split an input region into two or more output regions, by cutting 
the region with lines or loops. After each split, each region can be assigned
a different material, and surface displacement can be optionally applied.
In our framework, operators trace lines that are either contour lines of a smooth 
scalar field $f$ defined over $\mathcal M$, or integral curves of its gradient 
$\nabla f$.

\paragraph{Geodesic metric}
We define the scalar field $f$ upon geodesic distances, which depend on both the 
manifold $\mathcal M$, and a \emph{seed set} $S$ made of points and lines on 
$\mathcal M$ from which distances are computed. Much flexibility of our operators 
stems from the possibility to set seeds in a proper way, e.g., combining lines from 
the boundary of a region and points sampled inside a region or on its boundaries.

\begin{figure}[h!]
  \centering
  \includegraphics{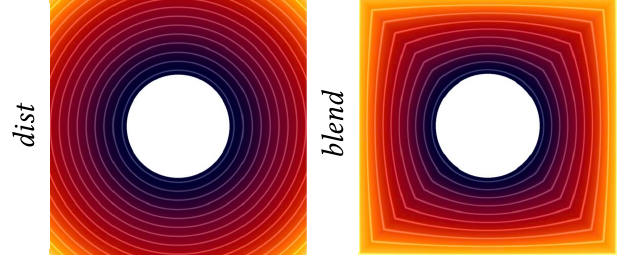}
  \caption{Field visualization: \emph{dist} field from the 
  center and \emph{blend} field from center to border.}
  \label{fig:field}
  \Description{Image}
\end{figure}

We consider two types of scalar fields, shown in \fig{fig:field}: 
the geodesic distance $dist_S(x)$ of surface points $x$ from the seed set $S$, 
and the blend between geodesic distance fields from two independent seed sets, 
defined as $$blend_{S, S'}(x) = \frac{dist_{S}(x)}{dist_{S}(x) + dist_{S'}(x)}.$$

The blend field, already used in \cite{Campen:2011} to derive parametrizations, 
has some nice properties: its values are always in the range 
$[0, 1]$, in particular $blend_{S, S'}(S) = 0$ and $blend_{S, S'}(S') = 1$; 
and it is anti-symmetric in the sense that 
$blend_{S, S'}(x) = 1 - blend_{S', S}(x).$
Moreover, if $S$ and $S'$ are lines, its contour lines are parallel to both 
of them in their vicinity, and the integral curves of its gradient meet them 
orthogonally.

\subsection{Operators}
\label{sub:operator_description}

In our framework, we use only four operators, which are summarized in 
\fig{fig:operators} and described in the following.

\paragraph{Contour operator}
The \emph{contour} operator traces the contour lines of the geodesic field, 
either \emph{dist} or \emph{blend}. 
The generated pattern depends on the selected seeds as well as on an isovalue $d$. 
When applied to a region, the contour operator generates two or more regions, 
splitting the surface with the contour lines at value $d$.
Depending on the selected seeds and the desired isovalue, this operator can 
be used to construct a variety of patterns. 
For example, when considering the distance from the region boundary, 
this operator produces outlines, while if used with points inside a region it 
produces polka dots.
By repeating the cuts at different equally spaced isovalues, the operator 
generates concentric or parallel stripes, where the density of stripes is 
controlled with a parameter.
The contour operator has been used to obtain most of the decorations on the 
teapot in \fig{fig:teaser} and overall it is the most frequently used operator 
in all our examples.

\paragraph{Stream operator}
The \emph{stream} operator cuts regions by tracing the integral curves of the 
gradient of the geodesic field, either $dist$ or $blend$. 
In this case, the seed sets are comprised of region boundaries, $S$ and $S'$, 
from which the integral lines emanate or end.
The generated pattern depends on the selected seeds and the desired number of lines, 
whose starting points are sampled uniformly along seed $S$. 
The stream operator generates $n$ stripes, each bounded by two consecutive 
integral curves. 
Such stripes stream outwards $S$, and towards $S'$ if the latter is specified.
The stream operator has been used to obtain the main pattern of the rug on the 
back of the elephant in \fig{fig:teaser}, as well as in several parts of the 
rolling teapot in \fig{fig:results}.
Contour and stream operators can be combined to generate grid patterns on 
cylindrical regions. 
See the example in \fig{fig:hierarchy} and the paragraph on 
\emph{Grouping and Hierarchy} below.

\paragraph{Voronoi operator}
The \emph{voronoi} operator partitions a region into the cells of a Voronoi 
diagram in the geodesic metric, each one corresponding to a point in the seed 
set. The user controls the seeds in the region, which can be either 
manually selected, or generated with Poisson sampling, as described in 
\sect{sub:graph}, to simulate the look of centroidal Voronoi tessellation.
Voronoi diagrams are used extensively in procedural pattern generation, 
but their common use is as a 2D or 3D texture. Our patterns are generated over 
the surface, so they have a more natural look. We use the Voronoi operator to 
obtain cellular-like patterns and as a drawing scaffold to place finer details.
See examples on the central mask and on the vase to the right 
in \fig{fig:procedural}, as well as on the fertility in \fig{fig:results}.

\paragraph{Polyline operator}
Finally, the \emph{polyline} operator draws a geodesic polyline on the surface 
by connecting a sequence of seed points.
The polyline splits a region into two if it is closed or it %is open but 
connects two points from the same boundary loop. 
If instead the polyline %is open and 
connects two different boundary loops, it does not split the region but %instead 
it is embedded into the region's boundary and changes its topological type.
For instance, two such lines can be used to cut a longitudinal strip from a 
cylindrical region, e.g., the spout of the teapot in \fig{fig:teaser}. 
The polyline operator is most often used to cut large, meaningful patches that 
are then refined and decorated with the other operators.

\subsection{Applying Operators}

\paragraph{Seed selection}
In our prototype, the user can select region boundaries, either completely or 
partially, as well as points on the surface.
A seed set is made of an arbitrary collection of such selections.
Points can be either selected individually, or sampled according to a Poisson 
distribution in the geodesic metric. Significantly different patterns stem 
from different selection as already shown in \fig{fig:operators}.

\begin{figure}[t]
\centering
\includegraphics[width=\linewidth]{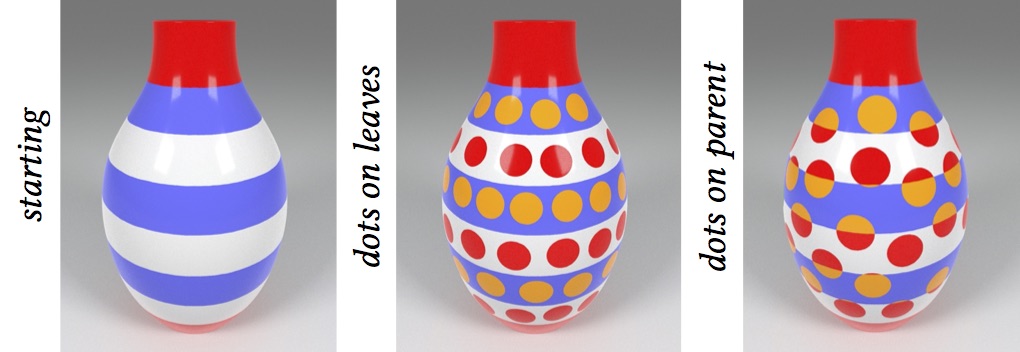}
\caption{(Left) The body of the vase has been partitioned into stripes 
with a  contour operator.
(Middle) A Polka dots pattern is obtained with a contour operator applied 
to Poisson point sampling in each stripe region separately.
(Right) The same pattern is applied on the parent region of the stripes, 
which is the body of the vase; 
in this case, the whole surface between the red strips is considered as a 
single piece for point sampling and distance computation. 
%(Lower Left) The grid pattern is obtained with a stream operator applied to 
%the parent region of the stripes in the Upper Left. 
%(Lower Middle) The white-blue stripes are obtained with a contour 
%operator applied between two opposite edges of the blue squares.
%(Lower Right) The grid pattern is obtained with a contour operator applied 
%between the other two edges in the parent regions of the stripes. 
}
\label{fig:hierarchy}
\Description{Image}
\end{figure}

\paragraph{Grouping and Hierarchy}
To further extend the capabilities of our framework, we support the application
of operators not only to the leaves of $\mathcal T$, but also to regions that 
are higher in the hierarchy. 
In this case, the operator defines fields and lines in the whole selected region, 
while all sub-regions that are descendant from it are sliced with such lines, 
and new leaf regions are generated accordingly. 
See examples in \fig{fig:hierarchy} and \fig{fig:macros} (grid).
This operation is natural in our framework, since field computation and 
split operators are independent. This same scenario required specialized cutting 
operations in the group grammars presented in \cite{Santoni:2016}. 
This extension also implies that we are less sensitive to the order in which 
operations are applied, a common concern in split grammars \cite{Schwarz:2015},
thus leaving more freedom to the user.
%which in turn means that users can edit more freely without needing to plan the 
%edits too much ahead. \fabio{mettiamo quest'ultimo pezzo dopo la virgola?}

\begin{figure}[t]
\centering
\includegraphics[width=\linewidth]{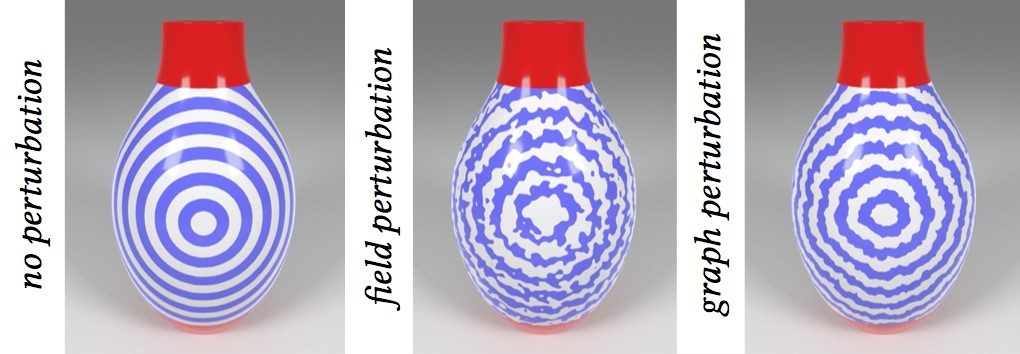}
\caption{The same pattern is generated without perturbation, with perturbation
of the scalar field and with perturbation applied to the surface before geodesic
field computation. 
Note how perturbing the surface instead of the resulting geodesic field preserves 
the topology of the generated regions, e.g., connected components.}
\label{fig:noise}
\Description{Image}
\end{figure}

\begin{figure}[t]
\centering
\includegraphics[width=\linewidth]{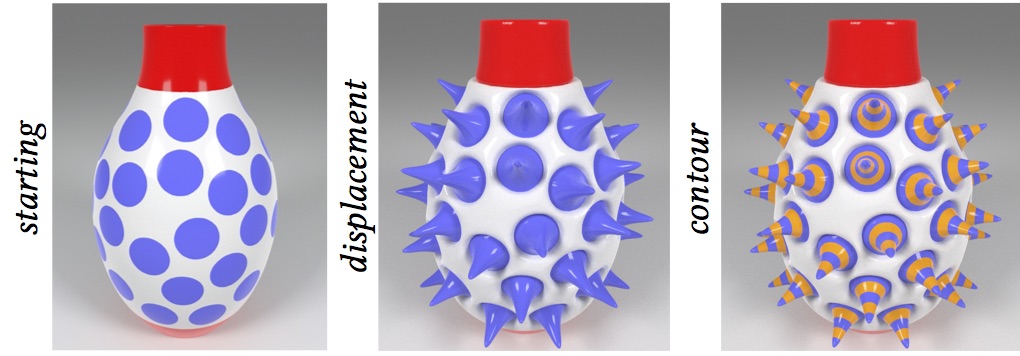}
\caption{
Example pattern generated applying strong surface displacement using different
profiles to create bulges and horn-like structures.
Note how we can coherently edit the horns with additional decorations
since distance fields are computed on the updated geometry.
}
\label{fig:displacement}
\Description{Image}
\end{figure}

\paragraph{Perturbation and displacement}
We can perturb geodesic fields and displace surface points 
to obtain a more ``handcrafted'', organic, look.
Scalar field perturbations simulate the imprecision of artists' hands;
while surface displacement simulates accumulation or removal of material. 
See examples in \fig{fig:noise} and \fig{fig:displacement}.

To simulate perturbation and handmade imprecisions, we apply 3D Perlin noise 
to the vertices of the mesh by offsetting their position along the normal 
direction before computing geodesic distances.
This perturbation is just symbolic and used only by the geodesic solver, 
while the actual mesh is not changed.
Geodesic distances computed on such warped geometry are therefore transformed 
accordingly to a coherent metric and produce irregular shapes, providing 
organic results. 
This works significantly better than perturbing the scalar field directly, 
since this can cause unwanted topological changes to the perturbed regions 
due to the fact that the perturbed field is no longer a distance field.
Perturbation can be controlled by setting the gain and frequency of the Perlin noise.
For instance, all the irregular blobs on the body of the teapot in 
\fig{fig:teaser} are actually circles from a perturbed metric; 
likewise, the irregular stripe patterns on the spherical bodies forming 
the lid of the same teapot are obtained with perturbation.
See also supplemental video.

\begin{figure*}[t]
\centering
\includegraphics[width=\textwidth]{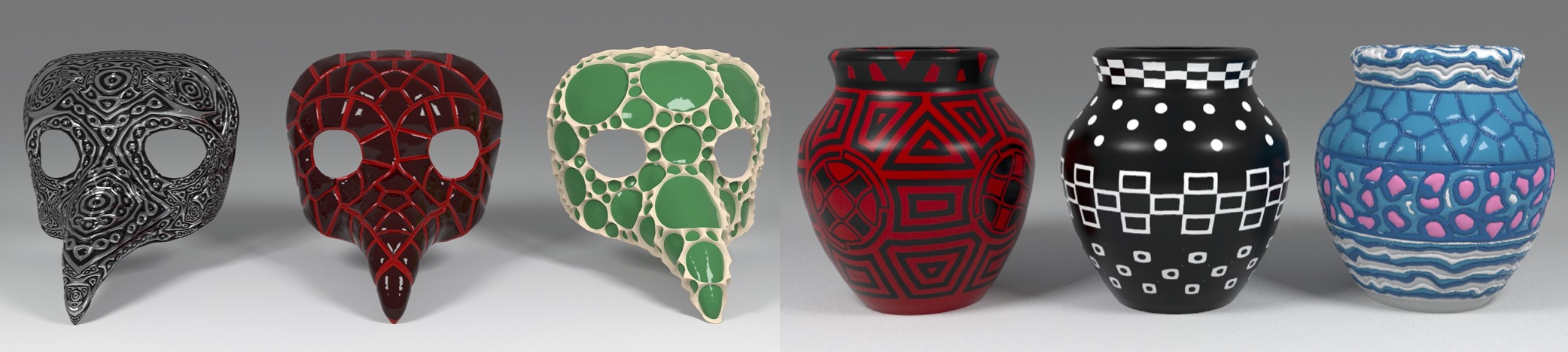}
\caption{Examples of fully procedural decorations created automatically (except colors).
Procedural patterns are created by recursively applying macros that are sampled stochastically. 
After each expansion, regions are regrouped as in \cite{Santoni:2016}, to achieve more structured results.
The composabilty of our operators guarantees that procedural generation is always robust 
and can be applied at any point of the editing session.
}
\label{fig:procedural}
\Description{Image}
\end{figure*}

Finally, we allow the user to apply arbitrarily-large surface displacement, 
after each operator is applied. The displacement is proportional to the geodesic 
distance from the region boundaries, being null at the boundaries to avoid 
discontinuities. Its profile is controlled by applying gain and bias operators 
from \cite{Schlick:1994}, to create bumps, pits, ridges or spikes.
This feature has been used extensively in most of our results, 
see Figures \ref{fig:teaser}, \ref{fig:procedural} and \ref{fig:results}.

\subsection{User Interface}
To create and edit decorative patterns, we have implemented a user interface 
that presents to the user the selection methods and split operators exactly 
as described previously.
We choose to use a straightforward mapping of the operators' types and parameters 
to the interface since we found it easy to manipulate them directly, 
without further remapping or interface tuning. 
To further simplify editing, we support hierarchy tree navigation and undo, 
respectively by maintaining the pattern hierarchy and by storing snapshots 
of the internal data structures.
\fig{fig:interface} shows the user interface that is also demonstrated in the 
supplemental video. 
We have used this interface to generate all results in this paper.

\begin{figure}[t]
\centering
\begin{tabular}{cc}
\includegraphics[width=\linewidth]{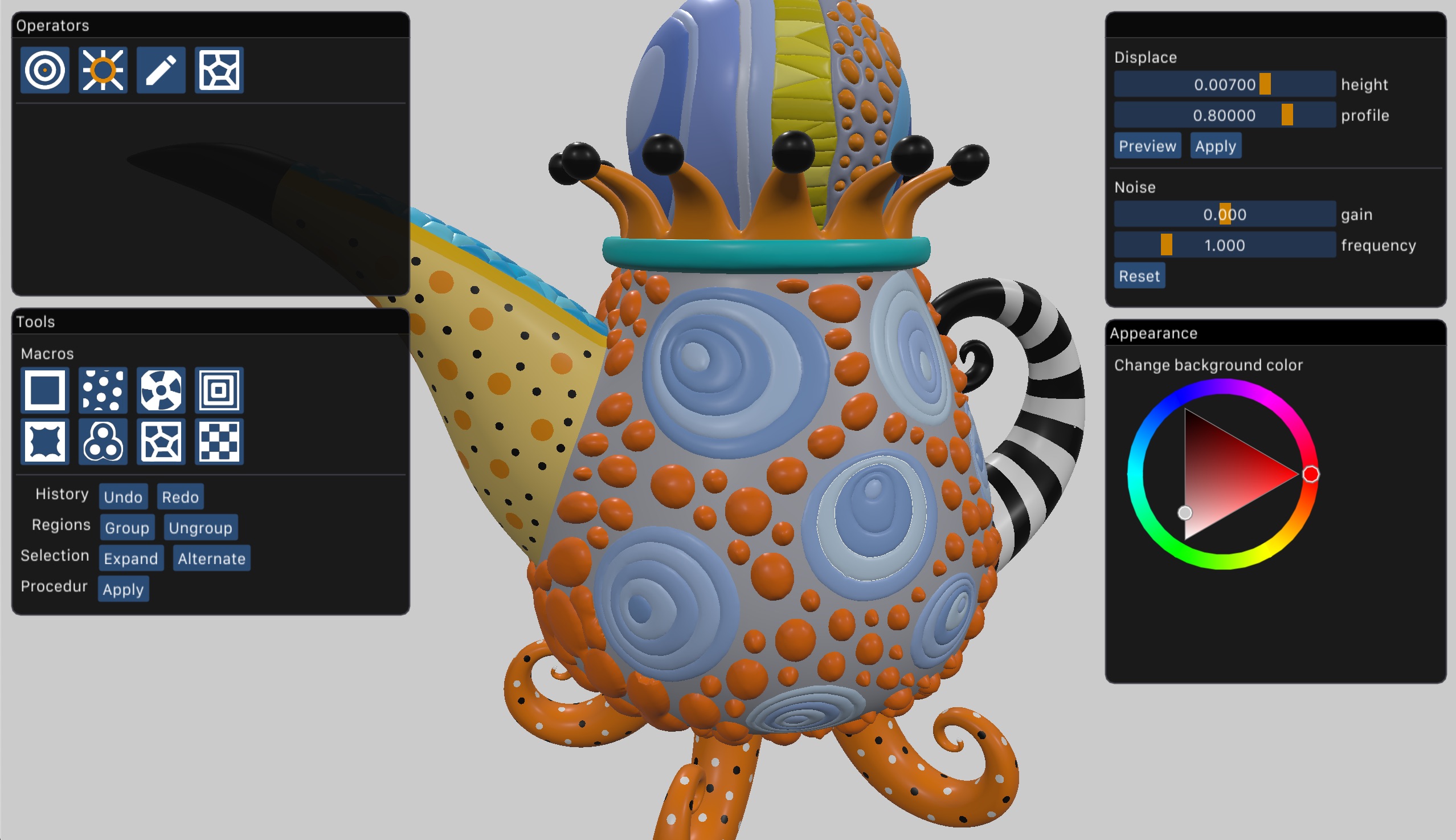}
\end{tabular}
\caption{
We implemented an application for real-time editing that let the user
decorate an input model applying the operators described in \sect{sub:operator_description} directly.
The user interface also features utilities for all the operations
described in \sect{sec:operators}, such as surface displacement,
perturbation, \emph{macros} and procedural pattern generation. }
\label{fig:interface}
\Description{Image}
\end{figure}
                        
\begin{figure}[t]
\centering
\includegraphics[width=\linewidth]{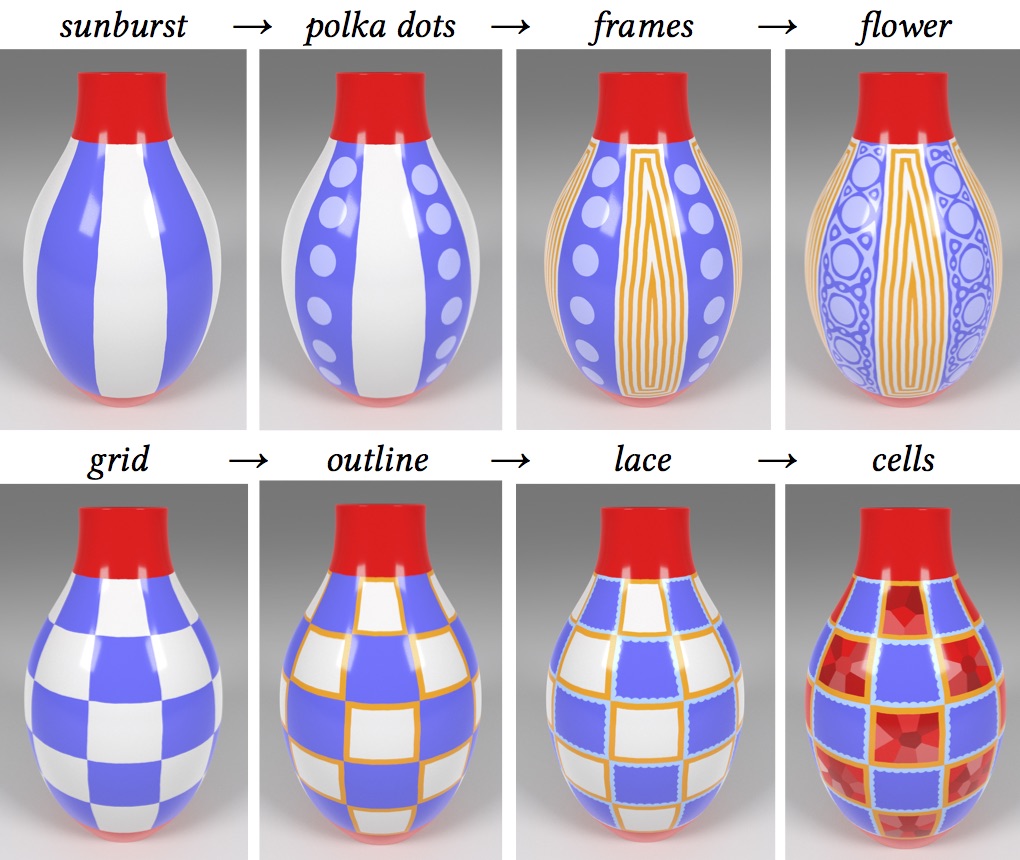}
\caption{
Example decoration created by applying four macros successively. 
(First row, left to right): Example decoration applying four macros successively.
(1) The \emph{sunburst} macro creates longitudinal stripes or wedges 
connecting the two borders of a cylindrical region, by applying the 
\emph{stream} operator and the two boundaries as seed sets.
(2) The \emph{polka dots} macro, in the blue region, creates dots by applying 
the \emph{contour} operator to a set of Poisson-sampled seed points from the 
selected region.
(3) The \emph{frames} macro generates concentric outlines by applying the 
\emph{contour} operator recursively to the boundary of the selected region, 
that in this case is the white one.
(4) The \emph{flower} macro creates flowing decorations by applying the 
\emph{contour} operator to the \emph{blend} between the distance from the 
region boundaries and the distance from Poisson-sampled points.
(Second row from the left)
Example decoration created by applying four macros successively.
(5) The \emph{grid} macro generates checkerboards by recursively applying the 
\emph{contour} and \emph{stream} operators between the opposite boundaries of a
cylindrical region.
(6) The \emph{outline} macro creates outlines by applying the \emph{contour} 
operator to all the boundaries off the selected boundaries.
(7) The \emph{lace} macro create embroidery-like decorations 
by applying the \emph{contour} operator to seed points uniformly sampled from 
the boundaries of the selected region.
(8) The \emph{cells} macro creates cellular patterns by applying the 
\emph{Voronoi} operator to Poisson-sampled seed points.
}
\label{fig:macros}
\Description{Image}
\end{figure}

\paragraph{Macros}
Overall we found that by combining selection, operators and hierarchy we can 
create very complex patterns with ease. 
We also found that some specific combinations of operators and seed sets 
were often chosen together to create distinctive and recognizable patterns.
To reduce manual work, we bundle these configurations into ``macros'', 
in a manner similar to 3D editors.
In our application, each macro is represented with just a button in the interface.
\fig{fig:macros} shows two simple sequences of editing using the eight 
macros we found most useful. See figure caption for a description.

\paragraph{Feedback Cycle}
Applying an operator always implies first computing a geodesic field
and then cutting the surface along its contour lines or integral curves. 
We show live previews of the final result before cutting the mesh, 
by just coloring the surface with real-time shaders.
This gives immediate, meaningful and precise feedback before actually 
modifying the model. 
In turn, this allows the user to work non-destructively while seeking good 
parameter values, and only commit the edit once it is good.

\subsection{Procedural Patterns}
The operators described before can also be applied automatically to 
generate fully procedural patterns. \cite{Santoni:2016} demonstrate how to 
generate 2D recursive patterns using group grammars, a variation of context-free 
grammars that can be used to describe stochastic recursive patterns. 
While our operators and selections are very different from theirs, 
the manner in which they can be recursively applied is similar.
Inspired by their work, we have implemented procedural patterns as described 
below using a formalism which is equivalent to a group grammar. In contrast to 
prior work, we integrate procedural generation in our interface, 
where procedural expansion can be applied to any region by just clicking a button. 
The resulting pattern can then be non-destructively edited interactively. 
The user controls procedural generation just by selecting the region to edit,
unlike \cite{Santoni:2016} that require users to edit grammars directly.
We follow a simpler approach since we focus on interactive editing and 
not grammar definitions, which are notoriously hard to edit, 
especially for designers. The resulting generator is equivalent to one example 
instance of group grammars, demonstrating that our operators and selections can 
also be controlled procedurally if desired.

\paragraph{Procedural Generation}
We generate procedural patterns by recursively applying macros 
with default parameters to the leaf regions of the pattern tree.
For each region, the macro to be applied is sampled stochastically, 
with a probability that depends on specific attributes of the region. 
Region attributes describe its overall shape as well as the way it was generated, 
and determine which macro would probably generate a pleasing result.
The selection tags and probabilities used for our patterns are included in 
the supplemental material. Here we describe the features we use to derive them.

For each region, we keep track of its boundary loops and corners, namely the 
intersection points where three or more regions meet.
For example, a region with one boundary and four corners will likely produce 
pleasing grid patterns. This is akin to \cite{Santoni:2016} where they instead 
track directionality of shapes, which is hard to do on surfaces since absolute 
tangent directions cannot be defined in a coherent manner.

As in group grammars, each region is assigned a type tag depending on the kind 
of operation that generated it and the number of region corners. 
For example, after an \emph{outline}, we differentiate regions between the 
inside and the border band, and after \emph{dots}, we differentiate the 
circle-shaped regions from the background. After each expansion, regions are 
grouped with the same rules used in \cite{Santoni:2016}, so that the same 
macro is then applied to each region of a group, to obtain more structured 
results. Some examples of such rules are to group regions by type, 
by attributes or with a cyclic rule to obtain alternating or checked patterns.
Procedural displacement is applied uniformly after the expansion is complete,
while the choice of colors and materials is left to the user.

\paragraph{Discussion}
\fig{fig:procedural} shows example patterns generated procedurally. These 
examples show that operators and selections can be controlled procedurally.
The main reason is that the operators can be applied to most regions 
independently of the region shape, i.e. they are closed under composition.
This means that we can apply them recursively at will, which in turn implies 
that we can use many procedural formulations introduced by prior work to control 
the generation of procedurals recursive details directly on surfaces.

%% file: 04_implementation.tex
% !TEX root = decosurf.tex

\section{Implementation}
\label{sec:implementation}

In the discrete setting we encode the manifold $\mathcal M$ as a finely 
tessellated triangle mesh $M$. We refine the mesh after applying each operator 
to precisely embed region boundaries as chains of edges, allowing us to 
represent regions exactly as groups of triangles.
Since we target design applications, we support real-time interaction on 
commodity hardware on meshes up to a few million triangles.

Our implementation relies on geodesic distance fields, which are notoriously 
expensive to compute, especially for large meshes. 
We use a simple graph-based geodesic solver that is efficient, scalable, 
accurate enough at our tessellation level, and easy to update as the mesh is 
refined and displaced. We achieve efficiency and scalability by encoding 
the mesh and the graph with compact and tightly coupled data structures. 
This data-oriented design has other benefits since it easily supports undos, 
serialization and rendering; we will not discuss these aspects in this paper, 
though.

In the remainder of this section, we discuss implementation details to aid 
in reproducibility. We will also release an open-source implementation upon 
paper acceptance.

\subsection{Representation}

\paragraph{Mesh data structure}
We encode triangle meshes with an indexed data structure augmented with face 
adjacencies, a.k.a. winged data structure \cite{Paol93}, compactly stored in 
three arrays: positions (\texttt{float[3]}), triangles (\texttt{int[3]}), and 
face adjacencies (\texttt{int[3]}). The indexed format provides a minimal 
representation of geometry and connectivity, while adjacencies provide support 
for efficient line tracing, region flooding and boundary computation. 

\paragraph{Pattern representation}
Our patterns partition the surface into regions.
We represent regions implicitly by labeling each triangle with the identifier
of the region containing it.
Hence, the collection of patterns covering the mesh is represented as an 
array of integers, one for each triangle.
We maintain the hierarchy of region tags in a tree data structure,
with a negligible space overhead, as it contains at most few thousands nodes 
even on our most complex results, as shown in \fig{tbl:results}.
Compared to storing a hierarchy of meshes, this separate representation 
is both significantly more compact and does not need to be updated as the 
mesh is refined.

\paragraph{Field representation}
Scalar fields are encoded at mesh vertices and extended by linear interpolation, 
while their gradient field is piecewise-constant on triangles. 
Operators compute either contour lines or integral curves, including geodesic 
paths, which cut the mesh along polylines. 
We split all triangles intersected by such polylines every time an operator 
is applied, so that regions always consist of discrete collections of triangles.

\subsection{Geodesic Computations}
\label{sub:graph}

\begin{figure}
\centering
\includegraphics[width=0.9\columnwidth]{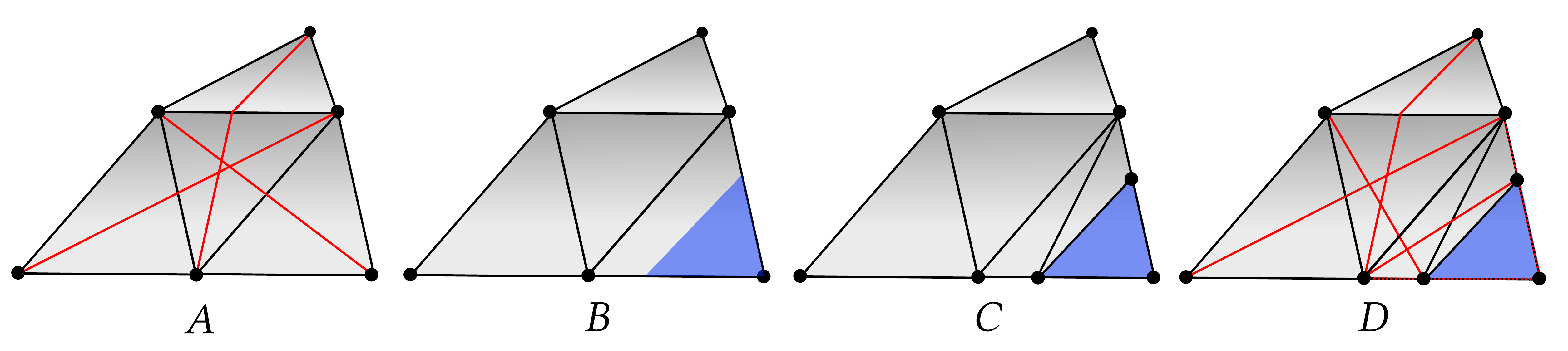}
\caption{
(A) Our graph has one node for each vertex in the mesh (black dots),
a bidirectional arc for each edge in the mesh (black lines) and for  
%a bidirectional arc that connects 
each pair of vertices that are opposite to an edge (red lines).
The weight of each arc is equal to the geodesic distance between the connected vertices.
%(B) The exact geodesic distance between two opposite vertices can be trivially computed by flattening 
%the pair of triangles that share the edge crossed by the geodesic path.
(B) When an operator is applied, the surface is partitioned into new regions (blue and grey).
The boundary that separates the new regions is a polyline crossing the edges of the
triangles in the mesh.
(C) The mesh is refined to embed the polyline, so that each triangle belongs to one region. 
(D) Only arcs in the graph traversing the split triangles are updated.
}
\label{fig:graph}
\Description{Image}
\end{figure}

\paragraph{Geodesic Graph}
Our solver is implemented using the graph exemplified in \fig{fig:graph}.
Nodes %of our graph 
correspond to mesh vertices, while arcs correspond to mesh 
edges as well as dual edges, i.e., arcs connecting pairs of vertices opposite 
to an edge. 
The length of each arc is computed as the exact geodesic distance 
between the vertices it connects. 
Arc lengths are stored in single precision to reduce memory usage, 
as our method is not prone to high error propagation
-- distances are just summed during graph visit -- 
while most approximation error stems from
discretizing geodesic paths along arcs of the graph. 
%An implementation in double precision provides negligible 
%improvement in accuracy at the cost of higher memory usage and worse performance.

We store the graph as adjacency lists with a simple array of arrays data structure, 
where we employ small vector optimizations for the adjacency lists.
This solution ensures that most of the graph is laid-out on a single 
contiguous chunk of data, which reduces heap pressure and improves cache 
locality during the graph visit.

We build the graph once at the beginning of the editing session, by using face 
adjacencies to construct dual edges, as shown in the supplemental pseudocode.
Then we \emph{locally} update the graph after each mesh refinement operation:
updates only involve nodes incident at split triangles, which are retrieved 
easily and efficiently, thanks to the implicit connection between vertices and 
nodes, and edges/adjacencies and arcs.
Updates upon displacement are also efficient since they only require 
recomputing edge lengths, without modifying the graph topology. 

\paragraph{Geodesic Solve}
We compute geodesic distance fields by wavefront expansion over the graph, 
shown as pseudocode in supplemental.
We adopt the small-label-first (SLF) and large-label-last (LLL) heuristics 
\cite{Bertsekas:1998}, which do not require a priority queue.
These techniques perform significantly faster than classical Dijkstra search 
on this kind of graphs \cite{Wang:2017}.
Our implementation aggressively exploits computation locality, 
by applying early exits when bounding the graph search to a region.
The general solver computes distance fields from any given set of vertices. 
Seed sets corresponding to lines are just sampled at their vertices.
Note that we use our solver just to compute the geodesic distance, 
while we do not trace geodesic paths with sequences of arcs in the graph, 
as the latter would result in wiggly polylines.
  
\paragraph{Poisson sampling} 
We use Poisson sampling to generate seed sets for various operators. 
We adopt a farthest point sampling technique \cite{Eldar:1997} under 
the geodesic metric. 
This scheme requires the computation of a distance field for each sampling point, 
which may become expensive as many points are sampled over large regions. 
We take advantage of the wavefront nature of graph search to significantly 
optimize the computation time.

We begin by computing the distance field from the region boundary. Then we 
iteratively select the vertex with maximum distance and add it to the seed set, 
and we update the distance field by expanding from the new seed, without 
resetting the already computed distances. 
This allows us to update the distance field only in the proximity of the 
new seed, by exploiting early exit when hitting nodes that already have a 
shorter distance from the previous set.
The search radius becomes increasingly smaller as points are added to the 
seed set, resulting in a significant speedup.
See statistics in the supplemental material.
This is another advantage of a graph-based solver over alternatives that are 
non-local.

\paragraph{Line tracing}
Operators %described below 
require extracting contour lines and integral curves, 
as well as cutting the mesh with such lines. 
Contour lines are extracted per triangle by linear interpolation. 
Integral curves and geodesic paths are computed by tracing the 
piecewise-constant gradient per triangle. 
Each triangle intersected by one such line is split along the corresponding 
segment, forming three new triangles.
When a triangle is split, arcs and nodes are added to the graph to represent 
new vertices and edges, and the adjacency of nodes in the split triangle is 
updated accordingly, as shown in \fig{fig:graph}.

\paragraph{Operators}
The operators described in \sect{sec:operators} are implemented on 
top of the functionality described before. 
The \emph{contour} operator requires one or two geodesic solves, for $dist$ and
$blend$ respectively, followed by the extraction of isolines and cuts along them. %application of marching triangles. 
The \emph{stream} operator requires the same solves, this time followed by 
cuts along integral curves.
In both cases, computation is bound to the region in which the operator is applied.

The \emph{polyline} operator is implemented by tracing geodesic paths between 
each pair of successive points. 
Each segment is computed starting at a vertex and ascending the gradient 
of the distance field generated from the previous vertex.
This operation requires one solve per segment: early exit occurs as soon as 
the target vertex is reached, so computation 
is bound to the intersection between the selected region and a geodesic 
circle having the segment as radius. 

The \emph{Voronoi} operator requires a solve for each element in the seed set,
which may become expensive for large seed sets. 
We again optimize this operator by exploiting early exits in graph search. 
We first compute the distance field from all seeds together to find its maximum value. 
We then set this distance as bound for early exit when computing the field for 
each seed. 
Intuitively, this ensures that each mesh vertex is visited roughly 
twice, so computation time is bounded regardless of the number of seeds. 
For each vertex of the region of interest, we collect the distance from its three closest 
seeds. 
Then we generate the Voronoi diagram by splitting all triangles that 
have vertices lying in different Voronoi regions, as in \cite{herholz2017diffusion}. 
See statistics in the supplemental material for time performance.

Finally, shape perturbation and displacement are trivially implemented by updating 
the edge lengths in the graph, and the positions of vertices, respectively.

\paragraph{Discussion}
Our system is very compact and coherent, because all operations rely on 
geodesic distance computation, as well as on few other straightforward operations: 
%i.e., 
line tracing and mesh cutting. 
Performance depends %mostly 
on our capability to make geodesic computations fast, 
and update our data structures efficiently. 
In this sense, the choice of our specific graph solver provides an optimal 
tradeoff under a variety of aspects, such as accuracy, speed, scalability, 
simplicity, dynamic update, and early exits.
Scalability, simplicity and ease of %dynamic 
update descend from using just 
the vertices of the mesh as nodes, and relations from local mesh topology as arcs. 
Nodes in our graph have average degree of 12, hence for a mesh with $N$ vertices
our graph has $N$ nodes and about $6N$ bidirectional arcs.

In contrast, graph-based methods using Steiner nodes \cite{Lanthier:1997bg,Lanthier:2001} 
are much more complex to maintain upon dynamic updates, and have a larger memory footprint. 
Their number of %bidirectional 
arcs increases quadratically with the number $k$ 
of Steiner nodes per edge: for $k=1$ there are $\sim$$4N$ nodes and $\sim$$18N$ arcs; 
while, for $k=3$, the count raises to $\sim$$10N$ nodes and $\sim$$84N$ arcs,
making these graphs impractical for large meshes even with moderately low values of $k$.

The DGG \cite{Wang:2017} and SVG \cite{Ying:2013} methods use graphs that do have 
just the vertices as nodes, but each node has a high degree (order $10^2-10^3$). 
Their memory footprint is high and they are slow to update upon 
mesh refinement. 
Finally, the method proposed in \cite{Campen:2013} relies just on the graph of edges, 
but computations to straighten paths are done on-the-fly, making the method suitable 
for point-to-point path computation only.

With the PDE method of \cite{Crane:2013}, each solve implies the solution of 
a sparse $N\times N$ linear system, which can be pre-factorized for a given mesh, 
making the solution fast at the price of pre-computation. 
The main concern in using this method is that every time we change the mesh 
topology or geometry the expensive factorization step needs to be repeated. 
See \sect{sub:performance} for a comparative analysis in terms of accuracy and performance. 

%% file: 05_results.tex
% !TEX root = decosurf.tex

\begin{figure*}[t]
\includegraphics[width=\textwidth]{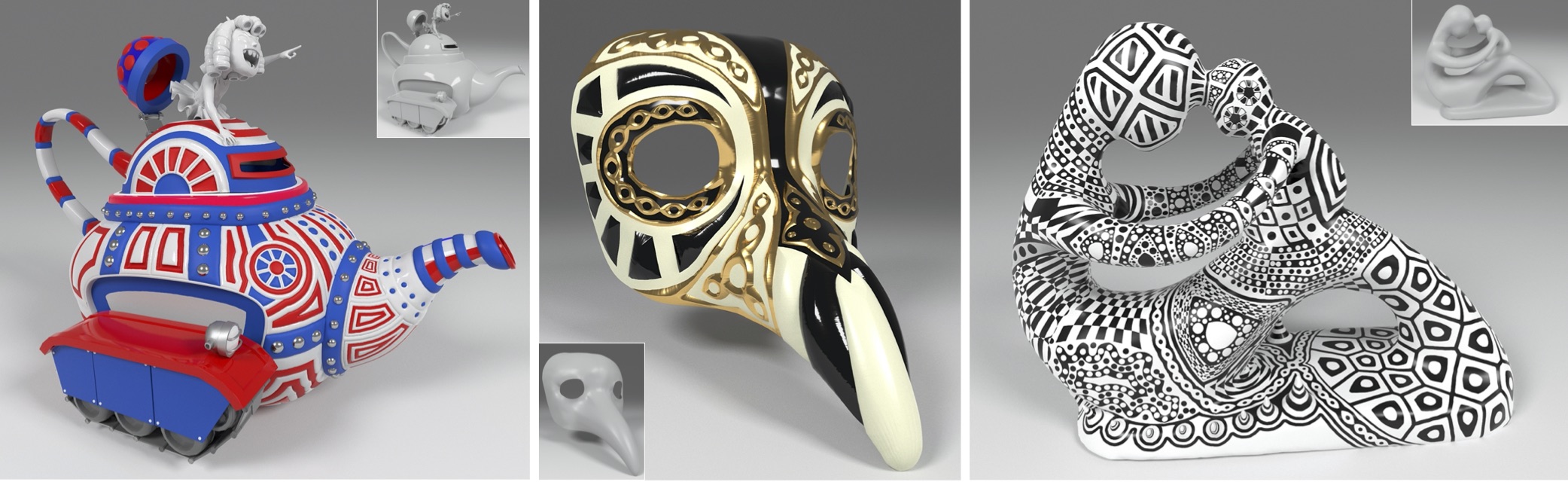}
\caption{Results created with our application starting from undecorated models, shown in the insets. Decorations were inspired by real world examples: the tank-teapot on the left
reproduces the playful look of handmade toys, the result in the middle matches the appearance
of carnival masks of Venice, the statue on the right is decorated with
intricate patterns that imitate tangles on ceramics.
Statistics about the models and the editing sessions are reported in \fig{tbl:results}
}
\label{fig:results}
\Description{Image}
\end{figure*}

\begin{figure*}[t]
\centering
\begin{tabular}{lrrrrrrr}
\toprule
model           & triangles & triangles & number of  & number of  & tree  & average time & memory  \\
name            & at start  & at end    & regions    & operations & depth & per operation & usage at end \\
\midrule      
fertility       & 0.50M    & 0.90M      & 1810      & 655     & 21  & 0.134s   &  96Mb \\
teapot          & 1.50M    & 1.65M      &  511      & 129     & 13  & 0.146s   & 173Mb \\
tank            & 1.45M    & 1.68M      &  636      & 145     & 10  & 0.189s   & 174Mb \\
mask            & 2.00M    & 2.15M      &   62      &  30     &  5  & 0.232s   & 228Mb \\
elephant        & 2.00M    & 2.61M      &  364      & 212     & 13  & 0.241s   & 278Mb \\
\bottomrule
\end{tabular}
\caption{Statistics on the editing sequences used for interactive decoration.
The average time per operation takes into account the time needed to compute the geodesic
field, cut the mesh, update the graph and the pattern representation data. }
\label{tbl:results}
\Description{Image}
\end{figure*}

\section{Results and Validation}
\label{sec:results}

We validated our pattern model in three manners. 
First, we modeled decorations that match real-world styles.
Second, we tested the accuracy and speed of the overall 
system and of the geodesic solver to show that it remains interactive.
Third, we validated our user interface with a user study to show that novices 
to the system can replicate given patterns.

\subsection{Editing Sequences}
\label{sub:models}

\fig{fig:teaser} and \fig{fig:results} show complex patterns created with our 
system during interactive sessions.
We chose to model patterns from real-world examples with different artistic styles 
to show that our model can capture intricate decorations made by artisans. 
\fig{tbl:results} shows statistics of the editing sequences corresponding to such images.

Overall we found that creating complex patterns is easy with our interface. 
We used from up to hundreds of single operations to create patterns made of up to 
1800 individual decorations, which correspond to regions in our model. 
Note that some such operations were applied as macros or procedurals, simplifying editing further.
The number of operations we employ is significantly smaller than using standard modeling tools 
with either polygonal modeling or sculpting workflows, e.g. see 
\cite{Denning:2011,Denning:2015} for statistics of common modeling sequences.

To gain a better sense of the recursive nature of the decorations, 
we report the depth of the pattern tree, which reached 21 in our most 
intricate result. This shows that by applying patterns recursively, 
even just a few times, we can greatly increase the complexity 
of the decoration while maintaining the editing manageable for users.

\newcommand{\centercol}[1]{#1}

\begin{figure*}[t]
\centering
\begin{tabular}{lr@{\hspace{0.2in}}r@{\hspace{0.2in}}rrr@{\hspace{0.2in}}rrrr}
\toprule
\multicolumn{2}{c}{model} & \multicolumn{1}{c}{\cite{Qin:2016}} & \multicolumn{3}{c}{\cite{Crane:2013}} & \multicolumn{4}{c}{ours} \\
\cmidrule(r){1-2} \cmidrule(r){3-3} \cmidrule(r){4-6} \cmidrule(r){7-10}
\multicolumn{1}{c}{name}       & \multicolumn{1}{c}{triangles} & \multicolumn{1}{c}{solve} & \multicolumn{1}{c}{build} & \multicolumn{1}{c}{solve} & \multicolumn{1}{c}{error} & \multicolumn{1}{c}{build} & \multicolumn{1}{c}{solve} & \multicolumn{1}{c}{error} & \multicolumn{1}{c}{update} \\
\midrule
kitten      & 300k &   3.0s & 1.49s & 0.068s & 0.5\% & 0.061s (24x) & 0.015s (4.5x) & 1.1\% & 0.013s \\ 
elephant    & 500k &  10.9s & 3.47s & 0.095s & 0.5\% & 0.098s (35x) & 0.027s (3.5x) & 1.1\% & 0.023s \\ 
fertility   & 500k &   3.5s & 2.86s & 0.123s & 0.5\% & 0.109s (26x) & 0.026s (4.8x) & 1.1\% & 0.023s \\ 
lucy        & 525k &   9.1s & 1.98s & 0.082s & 1.5\% & 0.175s (11x) & 0.034s (2.4x) & 1.6\% & 0.025s \\ 
mask        & 2.0M & 121.0s & 14.4s & 0.391s & 1.1\% & 0.585s (24x) & 0.114s (3.4x) & 1.2\% & 0.115s \\ 
nefertiti   & 2.0M &  20.3s & 14.8s & 0.345s & 0.9\% & 0.730s (20x) & 0.149s (2.3x) & 1.5\% & 0.139s \\ 
dragon      & 7.2M &  79.9s & 59.6s & 1.500s & 3.0\% & 2.613s (24x) & 0.446s (3.9x) & 1.5\% & 0.571s \\ 
% thai statue & 10M  & 111s  & 1.807s &  32\% & 5.059s (21x) & 0.835s (2.2x) & 1.7\% & 0.802s \\
\bottomrule
\end{tabular}
\caption{Comparison between our geodesic solver, and the solver from \cite{Crane:2013}. 
For the latter we use the implementation provided by the author, using Cholmod as backend. 
Columns \emph{build} report pre-processing times to pre-factor the system and to build the graph, respectively.
Build times for our solver also include the time to compute the triangle adjacency needed to build the graph.
Columns \emph{solve} report average time for computing the distance field from a single point source, 
where average is taken over 100 random samples.  
Column update reports the average time to update our graph after mesh split, where average is taken over
100 different long slices that roughly cut the mesh in half.
Root-mean-square errors are computed with respect to the exact polyhedral solution from \cite{Qin:2016}.
}
\label{tbl:comparison}
\Description{Image}
\end{figure*}

% \begin{figure*}
% \centering
% \includegraphics[width=\textwidth]{images/comparison.jpg}
% \caption{Comparison of geodesics computed with the correct method of 
% \cite{correctGeodesics} (left), the approximate method of \cite{Crane:2013} 
% (middle), and our approximate solver (right). 
% \fabio{Enrico: IS THIS REDUNDANT? DI BE DECIDED! CAN WE SKIP IT?}
% }
% \label{fig:comparison}
% \Description{Image}
% \end{figure*}

\subsection{Performance and Accuracy}
\label{sub:performance}

In our examples, we handle models between 500k and 2M triangles, 
which are further subdivided during editing.
Throughout the modeling sequences, our system remains interactive with 
computation times of about 0.2s per operation, including geodesic computation, 
mesh cutting and graph update. Memory usage is also compact, 
never exceeding 300Mb, which includes the mesh and the geodesic solver, 
as well as interface support data.
Performances were evaluated on a 2.9 GHz laptop with 16 GB of RAM running 
with a single-core for our application.

Fast geodesic computation is the main technical feature that enables us to 
model decorations well. We test the performance and accuracy of our solver 
by computing the geodesic field from a single source to all vertices on a 
variety of meshes, summarized in \fig{tbl:comparison}. 
While we use a very simple graph, our solver remains accurate enough with an 
error between 1.1\% and 1.6\% over a state of the art solver for exact 
polyhedral geodesics \cite{Qin:2016}. 
Computation times are between 0.015s and 0.446s for a single core implementation 
running on meshes between 300k and 7.2M triangles.
This speed is fast enough for all our modeling needs.

We compared our solver to a popular, approximate, geodesic solver based on 
PDEs \cite{Crane:2013} using the author's implementation.
The accuracy of our solver is just slightly lower on relatively small meshes, 
while getting better on large meshes. 
%, remaining consistently far below $2\%$ on all meshes. 
On the other hand, our solver consistently runs at roughly 
twice the speed, always remaining compatible with interaction, even when 
considering the additional times for update after mesh cutting.

In terms of pre-computation times, where \cite{Crane:2013} factorize a 
sparse matrix, while we build a graph, our method runs roughly between 
10 and 35 times faster. Although pre-computation is just performed once, 
the high times of \cite{Crane:2013} suggest that it would be hard to 
try a dynamic %factorization 
update after mesh edit.

We also compared our solver to a straightforward implementation of 
Lanthier's graph \cite{Lanthier:2001} with just one Steiner point per edge, 
using the same graph traversal algorithm.
In terms of accuracy, results are comparable to our solver.
We cannot objectively compare solve times as the two implementations 
were not equally optimized, but on average our solver was about 10 times faster.
Regardless of low-level optimizations, we assume our method to be more efficient,
as the Lanthier's graph with one Steiner node per edge is more than three times 
larger, as explained in \sect{sub:graph}.
Beyond accuracy and performance, our graph is much easier to maintain upon 
mesh cutting, and that was the determining factor, which made it a better choice 
for our application.

%Overall, we believe that our solver achieves the best trade-off between accuracy, 
%speed and ease of update to support the interactive generation of our patterns. 
We wish to remark that the choice of a geodesic solver is relatively orthogonal 
to our pattern model, though.

\subsection{User Study}
\label{sec:us}

We ran a user study to validate whether our prototype system is easy to use, 
whether it allows users to model complex patterns, and whether the style of 
the models we produce matches the style found in handmade artisanal objects.

\begin{figure}[t]
\includegraphics[width=0.95\columnwidth]{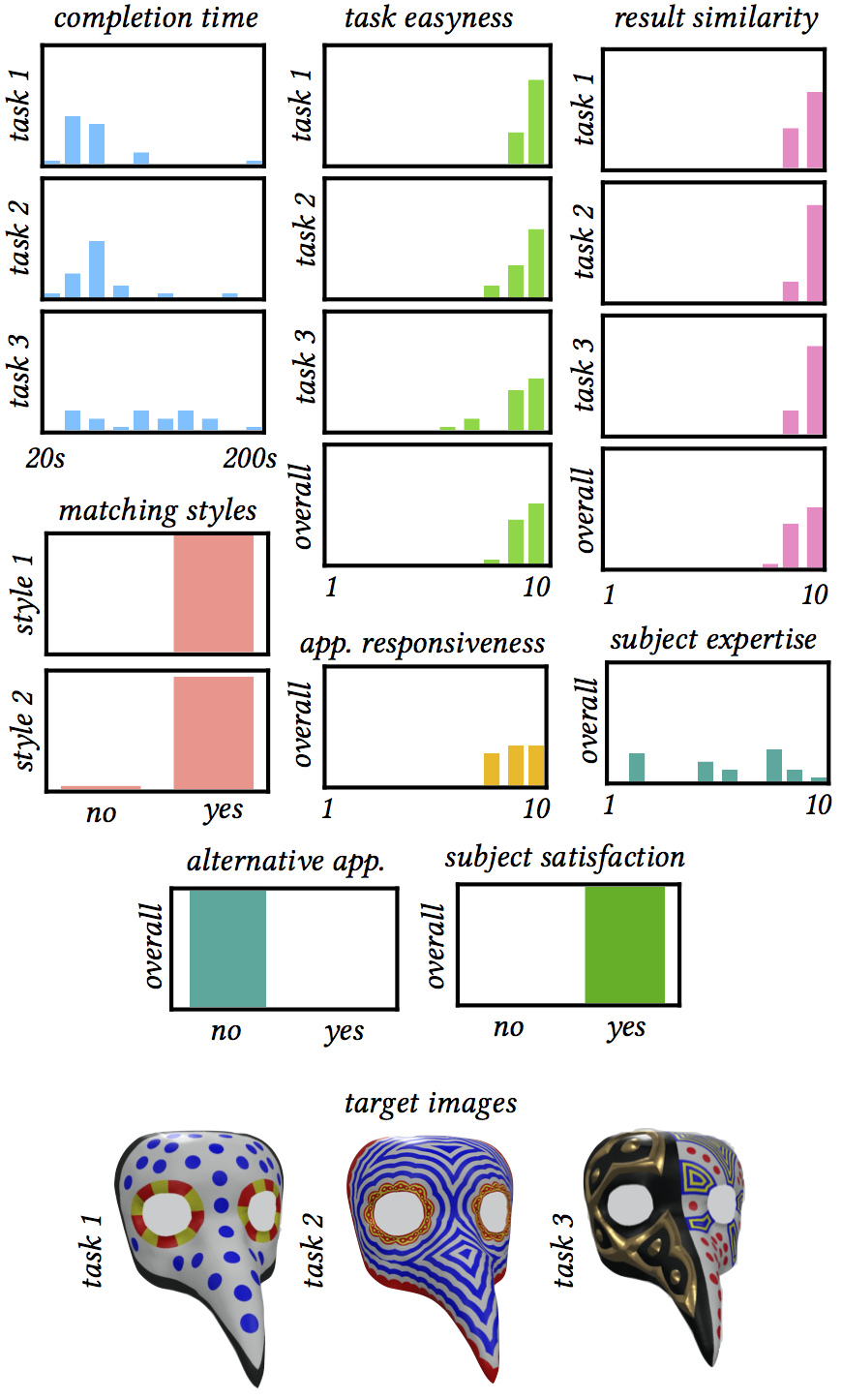}
\caption{Result data from the user study. Histograms on the left report
the time spent by the subjects to complete each task.
The bar chart just below shows that all subjects but one correctly recognized the
real world styles from which our results in \fig{fig:teaser} were inspired. 
The remaining bar charts show subjects' feedback about their tasks: 
users with different amount of expertise found the tasks easy to perform,
their results similar to the reference ones and the application responsive.
All subjects were satisfied with their editing experience with the application
and do not think that using another digital tool would allow them
to obtain the same results in the same time.
}
\label{fig:userstudy}
\Description{Image}
\end{figure}

\paragraph{Experimental procedure}
We asked 17 subjects with different degrees of expertise, ranging from novices 
to professional 3D artists, to use our prototype after a short training and
an unguided editing session on a model of their choice.
We asked subjects to perform three matching tasks of increasing
difficulty, in which they had to use the application to reproduce
a target image shown in a picture. 
Images %used in this study 
are provided in the additional material and in \fig{fig:userstudy}.

After each task, subjects were asked to rate the similarity of their result 
with the reference and to evaluate how easy they found it to complete the task, 
using a scale from 1 to 10.
We also asked subjects to rate whether they would have been able to obtain the 
same kind of results with a different 3D application, whether they found the 
interface responsive, and whether they were satisfied with the overall experience. 
Finally, we have shown to the subjects the teapot and the elephant decorated models 
from \fig{fig:teaser}, and asked them to recognize whether the style of the 
image matches real photographs of similar styles, compared to other alternatives. 
We include a copy of the final questionnaire in the supplemental material.

% \begin{figure}[t]
% \centering
% \begin{tabular}[@{}c@{\colspace}c@{\colspace}c@{}]
% \includegraphics[width=\iconsize]{images/512x512.png} &
% \includegraphics[width=\iconsize]{images/512x512.png} &
% \includegraphics[width=\iconsize]{images/512x512.png}
% \end{tabular}
% \caption{Target patterns to model in the user study.}
% \label{fig:targets}
% \Description{Image}
% \end{figure}

\paragraph{Quantitative Evaluation}
\fig{fig:userstudy} shows the results of our user study, where for all ratings 
we rejected the null hypothesis ($p \le 0.05$), i.e. those results are 
statistically significant.

All 17 users were able to successfully complete the reproduction tasks, spending
different amounts of time in the editing session, but never more than 4 minutes 
for each task, out of a maximum task length of 5 minutes.
All users rated their results quite similar, if not identical, to the reference ones. 
This suggests that our interface provides sufficient control to reproduce 
given complex patterns. All subjects also found the tasks easy to perform, 
and reported that they would not have been able to obtain the same kind of 
results with a different 3D editing software. 

In general, all subjects also found the interface responsive and 
%all of them reported that they 
were satisfied with the overall experience with the application,
confirming that our implementation remains interactive at all times.
All subjects but one correctly recognized the style from which our results were inspired, 
meaning that the models produced by our application match the look of real hand-crafted objects.

In conclusion, the user study demonstrated that users agree that our application is expressive, 
easy to use, and can produce results that match the look of real decorated objects.

\paragraph{Qualitative Feedback}
Some non-expert users informally reported that they were surprised by the 
complexity of the results they managed to obtain with the application and 
the ease with which they were able to control the editing operations.
We think that this cannot be explained only by the usability of the interface,
but rather it is a direct consequence of the intuitive design of our editing 
operators, which requires no expertise to be understood.
These facts suggest that the editing workflow of our application is well-suited
for non-technical artists and designers, too.

Experts users reported that they found the application responsive and pleasant to use.
We quote here some informal feedback we collected: ``The editing was surprisingly
fast and enjoyable. I did not have to think about triangles, edge loops or topology issues
as in Maya or Blender; I could just focus on the result.''

\subsection{Limitations}

\paragraph{Patterns}
While we introduce a general pattern model, recursive patterns based on tight 
packing of arbitrarily-shaped elements cannot be easily reproduced in terms 
of geodesics. 
While there is a large literature on this, we remind the reader that packing is 
NP-hard in general, so all methods proposed so far are necessarily strong approximations.

Another type of pattern that we did not specifically include is floral 
decorations. One possibility would be to adapt a region-growing model similar 
to the one used in procedural trees and express it in terms of geodesic paths.
The main concern though is that controlling floral arrangement over an arbitrary 
manifold remains hard since there is no global orientation to use while growing.
We leave the investigation of such further patterns to future work. 

%\paragraph{Geometry}
% Our current patterns are bounded just by isolines of a scalar 
% field, including geodesic circles, or streamlines of a vector field, 
% including geodesic lines, which play the role of the straight lines on a manifold. 
% Therefore, the we address just a domain of shapes equivalent to regions 
% in the plane that are bounded by polylines and circular arcs. 
% We are currently working on extending splines to manifolds, 
% in the context of the same computational framework, 
% in order to extend our domain further. 

\paragraph{Discretization}
We rely on fine tesselations to represent decorations and discretize 
geodesic computation. 
%Just like any other discretization, u
Using too few triangles
may introduce artifacts, such as wiggly polylines, due to mesh discretization, 
graph-based evaluation of distance fields and piecewise-constant approximation 
of gradient fields. While we did not witness these problems in our tests, 
they are nonetheless possible.
An alternative would be to separate the modeling and rendering phases. 
Since all our decorations can be encoded as operations on the manifold and 
its sub-regions, final patterns can be regenerated before rendering.
This would allow us to use an accurate representation of the manifold, 
such as a more refined mesh or a subdivision surface, more accurate geodesy, 
such as exact methods for meshes or PDE methods for subdivision surfaces 
\cite{DeGoes:2016}, and more accurate methods for isoline extraction and vector field 
tracing \cite{Kipfer:2003,Nielson:1999}.

%% file: 06_conclusions.tex
% !TEX root = decosurf.tex

\section{Concluding remarks}
\label{sec:conclusions}

We have presented an interactive method for generating recursive patterns on 
surfaces and its use to model real-world decorations. 
Our model consists of a closed algebra of regions, which can be split 
by applying four operators at will.
Operators are defined upon geodesic fields on the surface 
and our implementation relies on fast geodesic computations. 
A user study shows that the resulting application is effective, 
it is responsive on meshes in the order of one or a few million triangles, 
and it is easy to use for novices too.

Our system is easily extensible in a variety of ways.
Directional fields
%, which can be designed and controlled with consolidated methods 
\cite{Vaxman:2017} can be used to generate patterns based on 
further streamlines, which can be easily combed and 
constrained to lines and boundaries in the decor.
Fields can be also used to locally parametrize patches of the surface,
thus allowing for a mixed use of our vector graphics
together with raster graphics, i.e. textures, to create a composite decor.
Diffusion curves \cite{Orzan:2008} can be naturally incorporated too, to support 
smooth-shaded coloring of regions.